\documentclass{aa}
\usepackage[dvips,xdvi]{graphics}
\def\mlim{m_{\rm lim}}

\def \ie        {\hbox{\it i.e.,~}}
\def \eg        {\hbox{\it e.g.,~}}

\def \etal      {{\it et al.}}

\def \ssim      {\! \sim \!}
\def \ssimeq    {\! \simeq \!}

\def \snr       {{\it SNR}}
\def\\{\hfil\break}
\def\spose#1{\hbox to 0pt{#1\hss}}
\def\lta{\mathrel{\spose{\lower 3pt\hbox{$\mathchar"218$}}
     \raise 2.0pt\hbox{$\mathchar"13C$}}}
\def\gta{\mathrel{\spose{\lower 3pt\hbox{$\mathchar"218$}}
     \raise 2.0pt\hbox{$\mathchar"13E$}}}

\begin{document}
%
%
   \thesaurus{12         
              (12.07.1;  
               12.04.3;  
               11.03.1;   
               08.19.4)}  
   \title{Observing high-redshift Supernovae in lensed galaxies}

   \author{Tarun Deep Saini\inst{1} \and
           Somak Raychaudhury\inst{1}\and
           Yuri A. Shchekinov\inst{2,3}
       \thanks{saini, somak@iucaa.ernet.in; yus@rsuss1.rnd.runnet.ru}}

   \authorrunning{Saini, Raychaudhury \& Shchekinov}
   \titlerunning{Lensed high-$z$ Supernovae}
   \offprints{Tarun Deep Saini}

   \institute{Inter-University Centre for Astronomy and Astrophysics, 
               Post Bag 4, Ganeshkhind, Pun\'e 411007, India \and
         Department of Physics, 
               Rostov University, Sorge, 5, 344090 Rostov on Don, Russia \and
         Osservatorio Astrofisico di Arcetri, Largo E. Fermi, 5, 50125 
               Firenze, Italy
             }

   \date{Received ; accepted }

   \maketitle

   \begin{abstract}

Supernovae in distant galaxies that are gravitationally lensed by
foreground galaxy clusters make excellent cosmological candles for
measuring quantities like the density of the Universe in its various
components and the Hubble constant.  Distant supernovae will be more
easily detectable since foreground cluster lenses would magnify such
supernovae by up to 3--4 magnitudes. We show that in the case of the
lens cluster Abell~2218, the detectability of high-redshift supernovae
is significantly enhanced due to the lensing effects of the cluster.
Since lensed supernovae will remain point images even when their host
galaxies are stretched into arcs, the signal-to-noise ratio for their
observation will be further enhanced, typically by an order of
magnitude. We recommend monitoring well-modelled clusters with several
known arclets for the detection of cosmologically useful SNe around
$z=1$ and beyond.


   \end{abstract}

%
%

\section{Introduction}

Observations of distant sources with known absolute luminosity
(cosmological standard candles) are of primary importance to modern
cosmology, since the relation between the apparent magnitude,
luminosity and redshift of distant galaxies can be used to determine
the Hubble constant $H_0$, the deceleration (or density) parameter
$q_0$, and the cosmological constant $\Lambda$.  Observations of
standard candles beyond $z=0.05$ (where peculiar velocities are small)
can yield the value of $H_0$ with reasonably small uncertainty (\eg\
\cite{fil1,ham96}).  Observations of standard candles at $z\!>\!0.3$ are being
used for the determination of the fraction of the total energy of the
Universe in matter $\Omega_M$ and in some hitherto unknown form
$\Omega_\Lambda$ (\cite{riess98,perlmutter99,saini00}).  The study of
the gravitational magnification of standard candles at even higher
redshift will put tighter constraints on dark matter models of
cosmogony (\eg\ \cite{kb98, marri98, holz98, metcalf99, pm00}).

The work of Riess \etal\ (1998b) and Perlmutter \etal\ (1999) has
shown that, if detected significantly earlier than the epoch of their
peak luminosity, type~Ia supernovae (SNe Ia) would be the most useful
among cosmological candles at high redshift. However, the required
integration times for good photometry and for obtaining spectra of
such supernovae at redshift $z\ssim 1$ are estimated to be tens of
hours on a 10m telescope for $0\farcs 75$ seeing
(\cite{goop95}). These observations would clearly be more favourable if
these supernovae occur in galaxies magnified by gravitational lensing.

The magnification due to lensing can be significant enough to make
possible the detection of supernovae (SNe) in galaxies at high
redshifts ($z\!\gta\!1$). Narasimha \& Chitre (1988) first pointed out
that such events in giant luminous arcs (as in the A370 system) can be
used as a test of the lens models.  In the case of multiply imaged
supernovae, Kovner
\& Paczy\'nski (1988) deduce simple relations between the magnification
of such a SN, the separation of images, and the differences between
the arrival times of the event in different images.

Indeed, such SNe would serve as a unique probe for not only the
distribution of matter in the clusters, but also for studying the
source galaxies themselves. Due to the increased flux produced by the
magnification of the images, photometric and spectroscopic studies of
very distant galaxies can become possible.  This would enable us to
obtain information, which would be otherwise unavailable, about the
star formation process in the young galaxies (\cite{mel91,yee}), the
evolutionary status of AGN (\cite{sti}), and even the morphology of
distant galaxies (\cite{ctt}). Indeed, one of the farthest known
galaxies (at $z$=4.92, \cite{fra97}) would not have been
detected had it not been for the $10$-fold magnification by the
cluster CL1358+62 at $z=0.33$.

In this paper we address the feasibility of detecting lensed SN events
in high redshift galaxies which would be useful in the measurement of
cosmological parameters. From a qualitative point of view such a study
seems worthwhile for several reasons.  For a typical magnification of
3--4 mag (\cite{kovpac88}) the study of lensed SNe stretches the
usefulness of using them to characterize the distance ladder to further
distances by a factor of 4--6, or, equivalently, results in a
considerable decrease in the required duration of observation.
Furthermore, although galaxies lensed into arcs are resolved in one
direction due to stretching, a supernova in such a galaxy will remain
a point source, hence the signal-to-noise ratio (\snr) of a lensed
supernova in an arc will be superior to that of one in an unlensed
galaxy.  Finally, a cluster lens typically produces multiple images
with time delays between them being up to several months, thereby
making it possible to observe the same SN again, and measuring its
light curve more accurately, particularly in its pre-peak phase.

In the searches we propose, we do not have to be confined to one lensed
SN at a time. In many known cases of gravitational lensing of background
objects by galaxy clusters, several arcs and arclets can be found in
an area of the sky typically imaged by a single CCD frame.  In the case of
\object{Abell 2218}, for instance, there are 30 observed arclets
(\cite{ebb98, bez98}) with $R\le 23.5$ and $\mu_R\le 25.5$ between
$z=$0.5 to 1.5, so clearly a lot of galaxies can be simultaneously
monitored. This is also true of the cluster \object{Abell~2390}  at
$z=0.23$, in which, in addition to the famous ``straight arc'' (triple
image of a galaxy at $z=0.913$), there are at least 12 arclets ($R<21$)
between $z=$0.4--1.3 in an area of $2.7\times 2.7$ arcmin$^2$ around it
(\cite{bezecourt2390}). In the same area, in the magnitude range
$21<R<23.5$, there are four images of two galaxies at $z=4.05$
(\cite{pello2390,frye2390})  lensed by the same cluster.

Similarly, in a single HST WFPC field, which covers much less area
($\sim$5 arcmin$^2$) than most CCD cameras on terrestrial telescopes,
one finds 20 arclets brighter than $m_{\rm F675W}$=24 (corresponding
to at least 15 independent galaxies) beyond $z\!=\!0.7$ in the cluster
\object{Abell~370} ($z\!=\!0.37$, \cite{bezecourt370}), and 14 arclets
corresponding to 5 galaxies between $z=$0.5--1.5 in the cluster
\object{MS0440+0204} at $z\!=\!0.197$ (\cite{gio98}). Not all of
these, of course, would be magnified to the same degree, but on
average they would be magnified, making them easier to be observed
than in an unlensed case.

\begin{figure}[t]   
\centerline{
\resizebox{\hsize}{!}{\rotatebox{-90}{\includegraphics{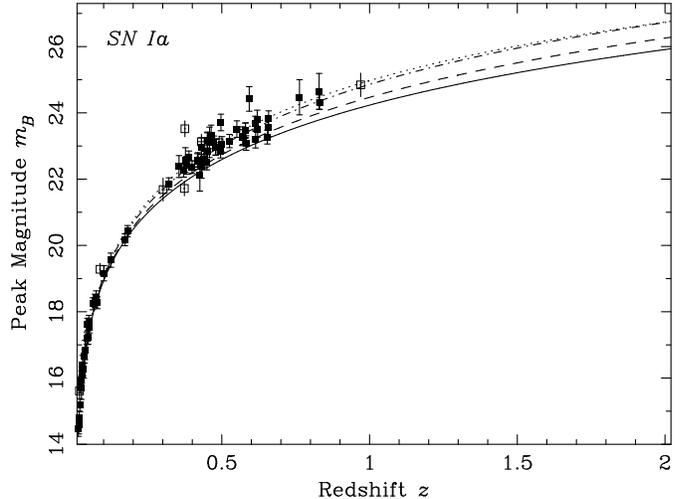}}}}
\caption[ ]{
The corrected apparent peak B~magnitudes of 79 published Type~Ia
supernovae (Perlmutter \etal\ 1999, Riess \etal\ 1998, Hamuy \etal\
1996) are plotted against redshift. The filled boxes represent
SNe that have been used elsewhere (\cite{perlmutter99,saini00}) 
to determine cosmological parameters, while the open
boxes indicate SNe that are generally left out (including SN1997ck at
$z=0.97$ which has not been spectroscopically confirmed to be
type~Ia).  The lines represent models with ($\Omega_M$,
$\Omega_\Lambda$): (a) Standard Cosmology (1.0, 0.0), full line, (b)
Open matter-only Universe (0.3, 0.0), dashed line (c) Perlmutter
\etal\ best flat model: (0.28, 0.72), dot-dashed line, and (d)
Perlmutter \etal\ best general model: (0.73, 1.32), dotted line.
\label{snplot}}
\end{figure}

Hereafter, we briefly review the usefulness of SN~Ia and SN~IIL in the
determination of cosmological parameters, and in \S3 quantify the
effect of a cluster lens on the detectability of high-$z$ SNe.  We
illustrate our case with the cluster lens Abell~2218.  In \S4, we show
that, in addition to this, for SNe detected in giant arcs, the
signal-to-noise ratio for the detection of SNe is enhanced by an order
of magnitude due to lensing. We summarize in \S5 in the light of
ongoing SN searches.

\section{Standard Bombs: Supernovae of Type~Ia \& IIL} 

Two subgroups of SNe seem to be  relevant for cosmological
use: SNe Ia and SNe IIL.
SNe~II are more frequent (by a factor of $\sim$4, \cite{vdB}) in late
spirals (Sbc-Sd), which are the most numerous among field galaxies.
It is more likely that supernovae at $z\!>\!1$, where the dependence
of the luminosity distance on models is the most sensitive
(Fig.~{\ref{snplot}}),
will be of type-II (\eg\
Madau \etal\ 1998).  Though SNe~II maxima in general are known to have
a large spread in luminosities, about half of them (those that have
``linear'' light curves, SNe IIL, \cite{yb,cap}) represent very good
standard bombs (\cite{gas}), since they have a small intrinsic scatter
($\sigma=0.3$ mag) around their peak magnitude
($\langle M_B\rangle=-17.05$, \cite{mib}, $H_0=75$).

However, SNe~Ia are much more luminous at their peak ($\langle
M_B\rangle=-18.95$ for $H_0=75$), with a smaller scatter in peak
magnitude, if corrected for the slope of their light curve
(\cite{perlmutter99, riess98}).

The major problem in the use of high redshift SNe as standard candles
lies in the identification of the type of the SN, and its photometric
calibration. The identification of the type of the SN depends much on
the shape of the light curve, which at high redshifts will be time
dilated, making it easier to determine its shape. The magnified flux
will render it easier to obtain a spectroscopic identification as
well.  It is therefore obvious that due to the considerable
magnification, lensed SNe will be far more suitable candidates than
unlensed SNe at the same redshift.

\section{The Effect of Lens Magnification on detectability}

\subsection{Gravitational magnification}

We summarize briefly the essential results needed for this paper.
Excellent reviews of gravitational lensing can be found
elsewhere (\eg\ \cite{bn,sch92}).

The basic equation which relates the angular coordinates of the source
($ \beta_1, \beta_2$) to those of the
image ($ \theta_1, \theta_2 $) is
\begin{equation}
{\vec{\beta} } =  {\vec{\theta} } - {\vec{\nabla} } \psi ( {\vec{\theta} }).
\end{equation}
The dimensionless relativistic lens potential satisfies the two
dimensional Poisson equation $ \nabla ^2 \psi ( \vec { \theta }) = 2
\kappa( \vec { \theta })$, the convergence $\kappa( \vec {
\theta }) = \Sigma ( \vec { \theta })/ \Sigma _{cr}$, $\Sigma ( \vec {
\theta })$ being the two dimensional surface mass density of the lens,
and $\Sigma _{cr}= (c^2/4 \pi G)(d_s/d_l d_{ls})$ is the critical
density. Here the distance between the observer and the source,
that between the observer and the lens and that
between the lens and the source are $d_s$, $d_l$ and $ d_{ls}$
respectively.

Gravitational lensing preserves the surface brightness of the light
rays. The flux of light received by an observer is directly
proportional to the solid angle subtended by the image at the
observer.  Since the solid angle of the image after lensing is, in
general, different from that of the source the observer can receive
more (or less) flux than in the unlensed situation. Thus galaxy
clusters can act as gravitational telescopes by collecting light from
the distant galaxies over a large area and sending it in our
direction.

The shape and size of the image are related to that of the source by
the transformation matrix $ M^{-1}_{ij} = \partial \beta_i/ \partial
\theta_j$. This matrix is generally written in the form
\begin{equation}
 M^{-1} =  \left( \begin{array}{cc}
1- \kappa -\gamma_1 & -\gamma_2 \\
-\gamma_2           & 1 - \kappa + \gamma_1
\end{array}
\right ),
\end{equation}
where $ \kappa $ is the usual convergence and $ \gamma_1 =
1/2(\partial ^2 \psi/ \partial \theta_1^2 - \partial ^2 \psi/ \partial
\theta_2^2), \gamma_2= \partial^2 \psi/\partial \theta_1 \partial
\theta_2 $ are the components of the shear. The magnification for a
point source is given by the Jacobian of the inverse mapping $ f :
\vec {\beta } \mapsto \vec{ \theta }$, which is, in general, one to
many.  From equation (2) we find that the magnification
\begin{equation}
\mu \equiv {\rm det}[M] = 1/((1-\kappa)^2 -\gamma^2),
\end{equation}
where $\gamma^2 = \gamma_1^2 + \gamma_2^2$, and is therefore different
for different images. The set of all the points where det$[M^{-1}]$
vanishes (singular points) in the source plane is called the caustic
set, and the images of the caustic set are called the critical
curves. A source of finite size (\ie\ a galaxy) close to the caustic
produces magnified images near the critical curves. Any point
source within these galaxies will also be substantially magnified.

For well-studied cluster lenses which have $\gta$10 arclets, the
model parameters of the lens can be constrained well enough so that
the magnification of a SN occurring at any point on an arclet can
be estimated reliably to an accuracy of $\lta 0.5$ mag. In this paper,
we suggest the monitoring of such well-modelled lenses to look for
cosmologically significant high-$z$ SNe.

\subsection{The enhancement of detectable events}

A SN event at a redshift $z \ssim 1$ is expected to have an apparent blue
magnitude $m_B \ssim 25$. If we are monitoring a system of arclets with an
observing setup (telescope and detector) of limiting magnitude of
detection $\mlim$ (given an acceptable value of S/N), then we would like
to estimate the probability that the SNe might be magnified by
an amount $\Delta m = m_B - \mlim$. Given this probability and the SNe
rates, one can obtain an estimate of the expected number of
detectable events.

Consider an area of the sky  of a few arcmin$^2$ around
the centre of a cluster at redshift $z_L$
being monitored with an array of CCDs.
Within this region, let the number of SNe occurring per year
in galaxies between redshifts $z_s$ and $z_s+dz_s$ be 
\begin{equation}
dN=N_0\, {\cal N}(z_s)\, dz_s
\end{equation}
where
\begin{displaymath}
\int^{z_{\rm max}}_{z_L}  {\cal N}(z_s)\, dz_s=1.
\end{displaymath} 

If $\mlim$ is the limiting magnitude, then the threshold
magnification at a  redshift $z_s$ for the source 
of unlensed magnitude $m_s$ to be 
detected will be 
$\mu_0(z_s) = 10^{(m_s- \mlim)/2.5}$.
The number of these 
SNe being magnified by a ratio $\mu>\mu_0(z_s)$ is
\begin{equation} 
dN^\prime= N_0\, {\cal N}(z_s) P(\mu>\mu_0 (z_s)) \, dz_s.
\label{eq:dnprime}
\end{equation}
where
\begin{eqnarray}
\lefteqn{ P\,(\mu>\mu_0 (z_s)) ={} } 
                                   \nonumber \\
 & & {}\ \ \ \ \frac{1}{\pi \beta_0^2}
          \int  \frac {\Theta[\,|\mu({\vec{\theta}})
         |-\mu_0(z_s)]\, \Theta[\beta^2_0 -\vec{\beta}(\vec{\theta})^2]}
           {|\mu({\vec{\theta}}) |} d^2 \theta, 
\label{eq:pmu}
\end{eqnarray}
for a source at redshift $z_s$.
Here $\Theta$ is the Heaviside step function, $\beta_0 (z_s)$ is the 
radius of the source (assumed circular) and the 
integral is performed over the field of the observed image.
Since a single source
can produce more than one image, the value of
this quantity, for a given threshold 
magnification $\mu_0$, can be greater than one.

Here we would quantify the enhancement in the detection of distant
SNe by 
\begin{equation}
\Phi_L(z)=\int_{z}^{z_{\rm max}}  {\cal N}(z_s) P(\mu>\mu_0 (z_s)) \, dz_s.
\label{eq:phil}
\end{equation}
This function represents the cumulative fraction of SNe that are
observed, given the limiting magnitude $m_{\rm lim}$ of the
observational setup, of the total number of SNe that occur between $z$
and $z_{\rm max}$ in the area of the sky that is being monitored.
This depends upon the number density and redshift distribution of the
host galaxies and the frequency of SN as a function of redshift.  This
should be compared with the quantity
\begin{equation}
\Phi_U(z)=\int_{z}^{z_{\rm max}}  {\cal N}(z_s) \Theta(\mlim-m(z_s)) \, dz_s
\label{eq:unphil}
\end{equation}
which is the corresponding fraction that would be observed in the
absence of the lens. For instance, since we assume the peak magnitude
of a Type~Ia SN to be $m_B=25$ at $z=1$, if $m_{\rm lim}=25$, the
value of $\Phi_U(z>1)$ will be zero whereas due to lensing $\Phi_L(z)$
can be finite till $z=z_{max}$.

\subsection{Example: the case of Abell 2218}

To estimate the typical fraction of SNe which would be seen behind a
cluster with a certain magnification, we consider the case of the
well-studied cluster lens Abell~2218 ($z=0.175$), for which good
published mass models exist. Here we use the model of Kneib \etal\
(1996), where the bimodal mass distribution is represented by two
cluster-scale clumps of dark matter centred on the two brightest
elliptical galaxies, their potentials modelled by the difference of
two pseudo-isothermal elliptical mass distributions (PIEMDs), with
an external truncation radius. In addition, the small-scale mass
structure is represented by galaxy-sized lenses corresponding to the
34 brightest galaxies belonging to the cluster, modelled by similar
functions with the appropriate parameters (velocity dispersion, core
radius and truncation radius) scaled to the observed luminosities of
the galaxies.

There are 258 background galaxies in the magnitude range $21.5<R<25$,
detected in the HST WFPC image of total area of 4.7 arcmin$^2$
analyzed by Kneib \etal.  Of these, 35 spectroscopic redshifts are
known, and another 18 redshifts are estimated in Ebbels
\etal\ (1998).  For other galaxies, random redshifts were chosen in
the range $0.175 \!<\! z<2.5$, from a distribution that conserves the
number of galaxies per unit comoving volume in a flat, matter
dominated universe.

The redshift dependence of the frequencies ${\cal N}(z_s)$ of both
SN~Ia and SN~II are taken from Madau \etal\ (1998), where the
evolution of cosmic supernova rates with redshift is computed from
estimates of the global history of star formation compiled from
multi-wavelength observations of faint galaxies. We assume that
redshift dependence of Type~IIL frequencies to be the same as that of
Type~II SNe taken as a whole.  Here the absolute value of the
frequency is not important, since we are interested in comparing the
number of SNe that would be detectable in the presence of a lens to
that if the lens were not there.

We take each of the background galaxies in our list, assuming their
intrinsic sizes to be 10 Kpc, and map them back to the source plane,
by means of the lens model and their measured/assumed redshift. The
integrals (\ref{eq:pmu}) is performed in the image plane by summing
over a fine grid, since the presence of the two $\Theta$ functions in
the integrand makes it difficult to evaluate them using Gaussian
quadrature.

\begin{figure}[tbh!]   
\centerline{\resizebox{\hsize}{!}{\rotatebox{-90}{\includegraphics{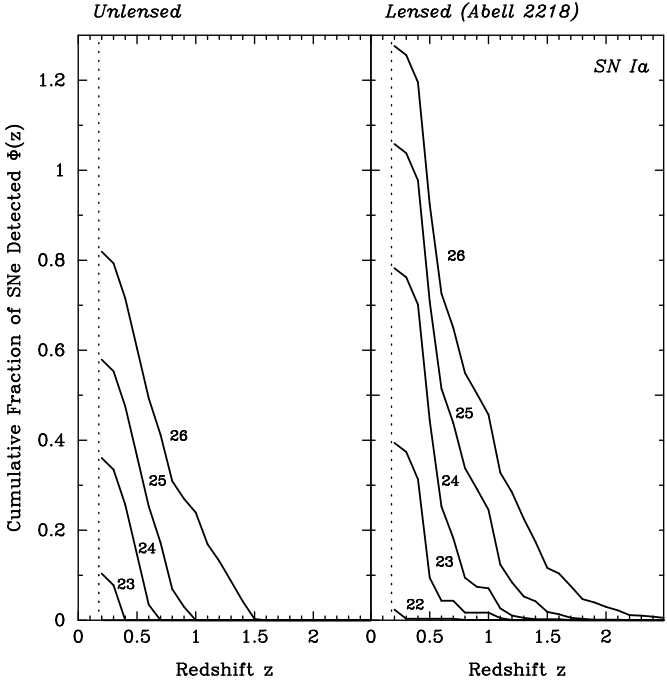}}}}
\caption[ ]{ The functions $\Phi$ (as defined in eqs~\ref{eq:phil}
\& \ref{eq:unphil}),
representing the cumulative fraction of detected SNe~Ia as a function of
redshift, is plotted for five different limiting B magnitudes of the
observational setup. 
The {\textit{right panel}} shows the function $\Phi_L$ (defined in
eq~\ref{eq:phil}) 
in the field of the lensing cluster A2218. Here we use the Kneib
\etal\ (1996) model of the lens, which comprises of a bimodal
distribution of dark matter and 34 of the brightest galaxies in the
cluster.  The {\textit{left panel}} shows the corresponding function
$\Phi_U$ (defined in eq~\ref{eq:unphil}), which represents the
function in the absence of the cluster lens, for the same values of
the limiting magnitudes.  The $z$-dependence 
of the frequency of SNe~Ia is taken from Madau \etal\ (1998).
The peak magnitude of SN~Ia at $z=1$ is
assumed to be $B=25$.  In the lensed case, $\Phi$ can be $>1$ since
sources can be mapped into multiple images.
\label{plotsn1}}
\end{figure}

\begin{figure}[tbh!]   
\centerline{\resizebox{\hsize}{!}{\rotatebox{-90}{\includegraphics{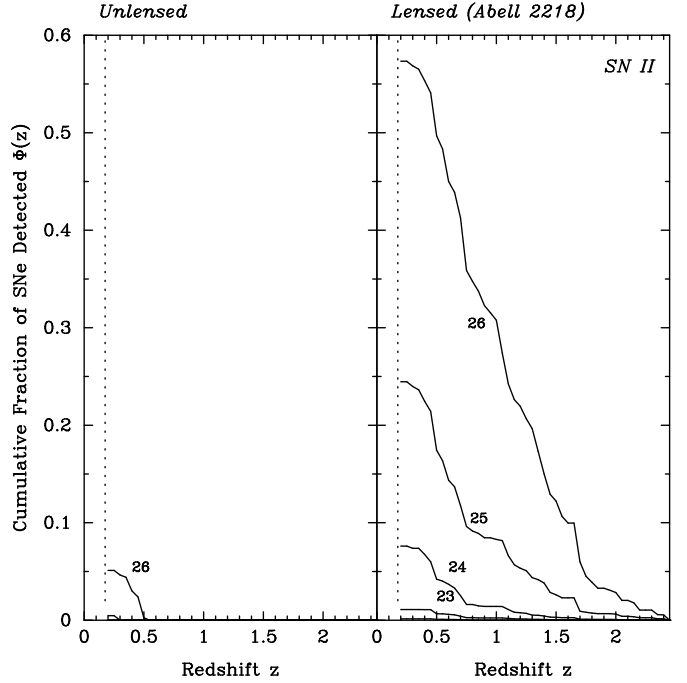}}}}
\caption[ ]{
The same as Figure~2, but for SN~IIL. The $z$-dependence 
of the frequency of SNe~II is taken from Madau \etal\ (1998)
and is assumed to be the same for SN~IIL. 
Here
the peak magnitude  at $z=1$ is assumed to be $B=27.5$.
\label{plotsn2}}
\end{figure}

We present the curves of $\Phi_L(z)$ for the Abell~2218 HST field for
both SN~Ia and SN~II in Figures~{\ref{plotsn1}} and \ref{plotsn2}
respectively, where we consider SNe in lensed galaxies in the redshift
range $0.175\!<z_s<\!2.5$. For comparison we also give the
corresponding values $\Phi_U(z)$ for the unlensed case, to show the
dramatic difference the presence of the lens makes. For example, for a
limiting magnitude of $B\!=\!25$, 20\% of all SNe beyond $z>1$ in the
field of the cluster A2218 will be detected, none of which would have
been detected if the lens were not present.

\section{Signal to noise enhancement in giant arcs}

One of the uncertainties in the accurate photometry of a SN
comes from the correction for emission from the host galaxy at the
site of the SN. The signal-to-noise ratio (\snr) related to this
uncertainty for SNe in lensed galaxies which form 
significantly elongated arcs or arclets
will be better than the
\snr\ in the unlensed galaxies. An enhanced \snr\ will in turn favour the
detection of a SN and the measurement of its characteristics.

For a seeing of 0.5--1 arcsec, a galaxy at $z=1$ would occupy
(referring to the area enclosing 90\% of the light) typically
$\sim$20 pixels. On the other hand, a supernova in the galaxy being a
point object would occupy the number of pixels covered by the seeing
disk, \ie\ 3--4 pixels.  Here we calculate the
enhancement factor $ \eta = \snr_{\rm lensed}/\snr_{\rm unlensed} $ for the
SN in the galaxy image.
	
For the sake of simplicity, we assume that in the unlensed case the
SN is not resolved. The total flux $F_{tot}$ that we would 
receive is the sum of flux from the SN and 
the galaxy, 
so the \snr\ for the detection of the unlensed SN is given by
\begin{equation}
 SNR \propto \frac{F_{tot} - F_{gal}}{\sqrt{{F_{tot}} + {F_{gal}}}}, 
\end{equation}
where $F_{gal}$ is measured long after the occurrence of the SN, so
that the latter no longer contributes to the flux from the galaxy.

In the lensed case the fluxes
should be multiplied by an appropriate
average magnification factor $\langle \mu \rangle$.  The lensed galaxy
is stretched into an arc in one direction, allowing us to define a
stretch factor $s$, which is the ratio of the angular size of the
seeing disk to that of the arc.  The amount of galaxy light
contaminating the SN flux is $s$ times the unlensed value.  Hence the
\snr\ enhancement $ \eta $ in the lensed case is given by

\begin{equation}
\eta = \sqrt{\langle \mu \rangle} \left(\frac{{\cal F}_{tot} - 
       sF_{gal}} {F_{tot} - F_{gal}}\right)
      \frac{\sqrt{ F_{tot} + F_{gal}} }
            {\sqrt{{\cal F}_{tot} +s {F}_{gal}}} 
\end{equation}
where $ {\cal F}_{tot}= s{ F}_{gal} + { F}_{SN}$.

The quantity in brackets is unity, since 
both numerator and the denominator are equal 
to { ${F}_{SN}$.
Denoting $r = F_{SN}/F_{gal}$ then the above formula simplifies to

\begin{equation}
\eta =  \sqrt{\langle \mu \rangle} \frac{\sqrt{r + 2}}{\sqrt{r + 2s}}. 
\end{equation}

For a typical case considered above, $r\ssim 0.2$, $s \ssim 0.1$,
${\langle \mu \rangle}\ssim 40$, which would give $\eta \ssimeq
15$. This shows that an order-of-magnitude enhancement in S/N ratio is
achievable in lensed SNe that appear in giant arcs.

\section{Conclusions}      

A considerable fraction of SNe in high redshift galaxies can be
magnified by foreground galaxy clusters.  This pushes the equivalent
observational distance to SNe further by a factor of 2--3, and thus
allows measurements of magnitudes and light-curves of high redshift
SNe in significantly shorter observational periods.

Even a small telescope with a limiting magnitude of $m_{\rm lim}=24$
will detect SNe~Ia up to $z\ssim 1.4$ in the field of a lens like
A2218, while in the absence of such a lens, the same setup would not
be able to detect SN~Ia beyond $z\!=\!0.7$ (Figure~2).  SNe of Type~II
will never be detected by such a setup beyond $z\!=\!0.5$ for $m_{\rm
lim}<26$, while in the presence of a A2218-like lens, they could be
detected up to $z\ssim 2$.

SNe occurring in giant arcs will have an additional gain of
signal-to-noise of $\sim$15, making it easier to observe them.  The
magnification for lensed SNe does not change dramatically for sources
between $z=1\!-\!2$, while the surface brightness of the host galaxy
decreases as $(1+z)^{-4}$, which in turn further improves the
signal-to-noise ratio.  Such SNe will also be multiply imaged, further
constraining mass models from time-delay measurements.

A number of projects have been searching for SNe at moderate and high
redshifts. Neither of the high-$z$ SN search projects
(\cite{perlmutter99, schmidt98}) has yet discovered a SN with a
favourable geometry behind a cluster. The low-$z$ Abell cluster search
(\cite{stromlo}) considers clusters out to $z=0.08$, again not ideal
for these searches.

In order to find high-$z$ SNe without having 
to wait for the commissioning of NGST, one
needs to continually monitor clusters with known arcs with a
favourable geometry (\eg\ Abell 2218, Abell 370, CL0024+17) in the
spirit of the SN Cosmology searches.  Once a SN event is
observed, one needs to calculate the magnification (weakly
model-dependent) from the location of the SN in the arc.  This will
yield the intrinsic apparent magnitude of the SN from the light curve,
which will enable us to calculate the distance to the supernova
independent of its redshift.  For a SN in the redshift range
$z=1\!-\!2$, this method can even yield a reliable value of $q_0$,
because we would be measuring $H_0$ in the relativistic regime. As
Fig.~{\ref{snplot}} shows, in this redshift range, SNe~Ia with
measured distances can distinguish, for example, between
$\Lambda=0$ and $\Lambda\ne0$ models even if the lens model
does not allow the estimation of the magnification to
better than 0.5 mag.

\begin{acknowledgements}
We thank J.-P. Kneib for sending us the data for his observation and
model of Abell~2218 in electronic form, and Rajaram Nityananda for
useful conversations.  YS acknowledges the hospitality of IUCAA, where
this work began, and the hospitality of Astronomisches Institut, Ruhr
Universit\" at Bochum, where he was at the final stages of the work. He
received partial financial support from the German Science Foundation
(DFG) within Sonderforschungsbereich, a travel grant from the IAU, and
a NATO Guest Fellowships grant (Ann. \#219.29) from the Italian
Consiglio Nazionale delle Ricerche. TDS thanks the University Grants
Commission, India for providing the financial support
(\#2-5/93(II)-E.U. II) without which this work would not have been
possible.  This research has made use of NASA's Astrophysics Data
System Abstract Service.
\end{acknowledgements}


\end{document}